\documentclass[journal]{IEEEtran}

\usepackage{makecell, longtable}

\usepackage{stfloats}

\usepackage[skip=1ex]{caption}

\usepackage{booktabs}
\usepackage{multirow}
\usepackage{csquotes}
\usepackage{amsmath,amssymb,mathtools}
\usepackage{bm}
\usepackage{cite}
\usepackage{url}

\begin{document}

\twocolumn[{
\copyright 2021 IEEE Personal use of this material is permitted. Permission from IEEE must be obtained for all other uses, in any current or future media, including reprinting/republishing this material for advertising or promotional purposes, creating new collective works, for resale or redistribution to servers or lists, or reuse of any copyrighted component of this work in other works.\\
	
This work has been accepted for publication in the journal IEEE Transaction of Cybernetics}]

%
\title{Gesture-based Human-Machine Interaction: Taxonomy, Problem Definition, and Analysis}
%
%
%

\author{Alessandro~Carfì, Fulvio~Mastrogiovanni
\thanks{A. Carfì and F. Mastrogiovanni are with TheEngineRoom,  Department of Informatics, Bioengineering, Robotics, and Systems Engineering, University of Genoa, Via Opera Pia 13, 16145, Genoa, Italy.}
\thanks{Manuscript received xx x, x; revised xx x, x.}}

\markboth{IEEE Transactions of Cybernetics  Class Files,~Vol.~x, No.~x, xx~x}%
{Shell \MakeLowercase{\textit{et al.}}: Bare Demo of IEEEtran.cls for IEEE Journals}
%



\maketitle

\begin{abstract}
The possibility for humans to interact with physical or virtual systems using gestures has been vastly explored by researchers and designers in the last twenty years to provide new and intuitive interaction modalities.
Unfortunately, the literature about gestural interaction is not homogeneous, and it is characterised by a lack of shared terminology.
This leads to fragmented results and makes it difficult for research activities to build on top of state-of-the-art results and approaches. 
The analysis in this paper aims at creating a common conceptual design framework to enforce development efforts in gesture-based human-machine interaction. 
The main contributions of the paper can be summarised as follows: 
(i) we provide a broad definition for the notion of functional \textit{gesture} in human-machine interaction, 
(ii) we design a flexible and expandable gesture taxonomy, and 
(iii) we put forward a detailed problem statement for gesture-based human-machine interaction. 
Finally, to support our main contribution, the paper presents, and analyses $83$ most pertinent articles classified on the basis of our taxonomy and problem statement.
\end{abstract}


%
\IEEEpeerreviewmaketitle

\section{Introduction}  
\label{sec:introduction}

Gestures have been explored as a communication channel both in human-computer (HCI) and human-robot (HRI) interaction. In these domains, usually, gestures unlock a new communication channel to send \textit{intentional} commands to a machine. We refer to this kind of interaction as \textit{functional} human-machine interaction (HMI), and we distinguish it from the notion of \textit{social} HMI\footnote{For the analysis considered in this paper, we argue that there are many common aspects for what regards HCI and HRI, and therefore we will refer in the paper to the broader notion of HMI.}, according to which a machine encodes and exploits human cognitive models for a better interaction.
Consequently, we refer to gestures employed in functional HMI as \textit{functional gestures}.

In this work, we will focus on functional HMI, and we will critically analyse the widely accepted idea that, in these domains, gesture-based interaction is a more natural alternative to classic tools such as the keyboard, mouse, or joystick.
Gesture-based interaction is of course \textit{natural} for social HMI, as well as in functional HMI scenarios in which each human motion can be piece-wise mapped to a given machine status, e.g., the manipulation of objects in virtual reality by tracking the human hand, where the virtual object pose can be mapped to a Cartesian reference frame centred on the hand itself.
Nevertheless, when we define a synthetic gestural language to substitute such tools as the keyboard or the mouse, we are not creating a natural interaction modality \cite{Cassell2010}, since we are not defining a new interaction language, rather we superimpose the gestural language to the one entailed by the keyboard or mouse.
To this extent, gesture-based interaction provides an alternative that in certain scenarios can yield a better user experience. 
For example, in automotive, gestures can be used to interact with the infotainment system without taking the eyes off the road \cite{Buddhikot2018}. 
Therefore, the overall idea in developing gesture-based interfaces should be to provide a new easy-to-use tool able to substitute or at least integrate the classic ones. 

In the literature, gesture-based interaction with machines has gained much attention in the past few years in the context of the Industry 4.0 paradigm, whereby inherently safe, task-adaptive, and easy-to-program collaborative robots are expected to work alongside and cooperate with human operators in shop-floor or warehouse environments \cite{kagermann2013recommendations}.
The need for a shared workspace where human operators and robots can perform turn-taking or joint operations in a safe and effective way has steered research in human-robot collaboration (HRC) along different directions.
Whilst safety aspects in HRC have been predominant in research, and have been grounded by the use of different sensing modalities, e.g., force/torque sensors \cite{haddadin2008collision}, touch sensors \cite{denei2015towards}\cite{albini2017enabling} and vision \cite{lasota2014toward}, also issues related to human-robot task allocation \cite{darvish2018flexible}\cite{capitanelli2018manipulation}, and robot behaviour programming \cite{billard2008robot} have been investigated.
New alternatives have been explored as well to enrich the interaction process, leveraging speech \cite{norberto2005robot}, touch screens \cite{singh2013interface} and human gestures.

\begin{table}[t!]
    \caption[Literature summary for gestural HMI]{Summary of gestural usage for human-robot interaction in the literature.}
    \centering
    \label{table:stateofart}
    \begin{tabular}{c|c|c|c}
    \multirow{2}{*}{\textit{Article}} & \multirow{2}{*}{\textit{Sensor}} & \textit{Number of} & \multirow{2}{*}{\textit{Description}} \\  
     & & \textit{gestures} & \\
    \hline
    \rule{0pt}{2.6ex}
    \cite{Kortenkamp1996}  & Stereo camera & 6 & Static arm poses \\ 

    \multirow{2}{*}{\cite{Kuno2000}}  & \multirow{2}{*}{Camera} & \multirow{2}{*}{-} & Discrete hand and \\
     & & & fingers motions \\

    \multirow{2}{*}{\cite{Hasanuzzaman2004}}  & \multirow{2}{*}{RGB camera} & \multirow{2}{*}{8} & Bimanual static hand\\
     & & & and fingers poses \\

    \multirow{2}{*}{\cite{Neto2009}}  & \multirow{2}{*}{Accelerometer} & \multirow{2}{*}{12} & Discrete arm\\
     & & & motions \\

    \multirow{2}{*}{\cite{Xing-HanWu2010}}  & \multirow{2}{*}{Accelerometer} & \multirow{2}{*}{6} & Discrete arm \\
     & & & motions \\

    \multirow{2}{*}{\cite{Iengo2014}}  & \multirow{2}{*}{RGB-D camera} & \multirow{2}{*}{5} & Discrete arm \\
     & & & motions \\

    \multirow{2}{*}{\cite{Cicirelli2015}}  & \multirow{2}{*}{RGB-D camera} & \multirow{2}{*}{10} & Discrete arm \\
     & & & motions \\

    \multirow{1}{*}{\cite{Lai2016}}  & RGB-D camera & 1 & Static hand pose\\

    \multirow{2}{*}{\cite{Coronado2017}}  & \multirow{2}{*}{Accelerometer} & \multirow{2}{*}{-} & Continuous arm \\
     & & & motion \\

    \multirow{2}{*}{\cite{Bolano2018}}  & \multirow{2}{*}{RGB-D camera} & \multirow{2}{*}{-} & Continuous hand \\
     & & & motion \\

    \multirow{2}{*}{\cite{Islam2019}}  & \multirow{2}{*}{RGB camera} & \multirow{2}{*}{8} & Static hand and \\
     & & & fingers poses \\
    \end{tabular}
\end{table}

Functional human-robot gestural interaction has been explored since late 1990s when the recognition of six different arm gestures has been used to control a wheeled robot \cite{Kortenkamp1996}.
Gesture-based interaction has been traditionally paired with speech-based interaction \cite{Bolt1980, Rogalla2002} to enrich the communication spectrum, and it can substitute speech entirely in noisy environments \cite{Kortenkamp1996}, or used to substitute such tools as teach pendants when human operators can not get their hands free.
Since first attempts, gesture recognition has been used in different applications to interact with robots. 
User-defined hand and finger gestures have been coupled together with face identification to allow people with disabilities to control an intelligent wheelchair \cite{Kuno2000}. 
Similarly, head orientation and bi-manual gestures perceived using an RGB camera have been selected to communicate with the pet-like robot AIBO\footnote{Web: \url{https://us.aibo.com/}} \cite{Hasanuzzaman2004}. 
Alternatively, wrist-worn accelerometers have been used to detect arm gestures aimed at controlling a 6-DoF manipulator \cite{Neto2009} and providing commands to wheeled robots \cite{Xing-HanWu2010}. 
Arm gestures have been explored to provide task level information to a mobile manipulator \cite{Iengo2014}, and to control a wheeled robot \cite{Cicirelli2015}. 
While pointing gestures have proved useful in indicating directions to assist a humanoid robot in navigation tasks \cite{Lai2016}, the usage of gestures to specify commands has been explored in difficult scenarios where other kinds of communication are very likely to fail, such as underwater \cite{Islam2019}. 
The communication of a discrete command to an intelligent system implies the usage of a \textit{discrete} gesture, while continuous commands used to tele-operate a wheeled robot \cite{Coronado2017} or a manipulator \cite{Bolano2018} necessitate \textit{continuous} gestures\footnote{This is an intuitive introduction to the concept of discrete and continuous gestures that will be formalised in the next Section.}.

Table \ref{table:stateofart} summarises the previously discussed papers, describing the used sensors, the number of gestures (if applicable), and the gesture types. In the Table, we could specify the number of gestures when the referenced paper was considering discrete gestures and not continuous ones since discrete gestures imply a gesture \textit{dictionary}. HRC is only one of the possible examples whereby gesture-based interaction can be exploited. But, it constitutes a compelling use case because it entails a physical system with embodied reactions to gestures. It is noteworthy that we consider HRC as a motivating scenario, and our analysis and conclusions are by no means limited to it. The authors of this paper extracted the gesture description outlined in the Table from the content of the original papers. However, this process can be difficult since, usually, an accurate gesture description lacks. This lack is a symptom of a deeper problem in this research field, i.e., an almost complete lack of standards and a concrete difficulty in building on top of the current state-of-the-art. This problem is even more acute in HCI scenarios. Vuletic et al. \cite{Vuletic2019} highlighted, as well, the difficulties faced by the scientific community to build on top of existing work because of a lack of public datasets and a clear problem statement in gesture-based HMI research. Although it provides an excellent contribution in reviewing gesture-based HMI approaches, nonetheless, Vuletic et al. \cite{Vuletic2019} does not provide any new common framework to be shared by the whole community and therefore adopted for future research.


To tackle these problems current literature lacks: 
(i) a common agreement of what a gesture is, which we aim at addressing in the context of functional HMI interfaces and not limited to HRI;
(ii) a comprehensive non-hierarchical taxonomy describing functional gestures in their relations with a generic intelligent system, and we provide accordingly a new gesture taxonomy based on a reasoned analysis of existing ones; 
(iii) a clear analysis of the involved problems in using gestures for functional HMI, which we better formalise as the \textit{Gestural Interaction for User Interfaces} (GI-UI) problem.   
As a consequence, the paper focuses only on intentional gestures and it is organised as follows.
Section \ref{sec:ui_gestures} provides an initial background and aims at laying the ground with the definition of what a gesture is.
Section \ref{sec:2:taxonomy} introduces a multi-modal taxonomy of gestures.
Section \ref{sec:problem_statement} discusses the problem as entailed by gesture-based HMI.
The role of human factors in gesture-based HMI is analysed in Section \ref{sec:humanfactor}.
A non-exhaustive classification of existing literature, exploiting all the conceptual tools developed in the paper, is presented in Section \ref{sec:classification} to support our main contributions. Section \ref{sec:eou} presents an example of usage for the proposed taxonomy and problem statement. 
A discussion and the conclusions follow.


\section{User Interfaces and Gestures}
\label{sec:ui_gestures}

\subsection{Background}
\label{sec:background}

In HMI, the \textit{user interface} is a system, composed either by physical or software components, which allows someone to use a machine or an intelligent system.
Obviously enough, the kinds of such machines or intelligent systems someone can interact with are countless, and vary from such physical systems as robots, to disembodied software applications. 
In interfaces which we can term as \textit{classical}, user interaction is mediated by physical tools, e.g., a keyboard, and usually feedback is provided as a reasoned combination or sequence of visual stimuli, e.g., in the case of a graphical user interface (GUI).

In this paper, we focus on functional gesture-based interaction for user interfaces, and in particular, we narrow down our attention on its technological requirements. 
In GUIs, a user directly interacts with the machine or intelligent system by means of gestures, although a physical device is still needed to perceive the gesture itself. 
As far as feedback is concerned, the classical approach could be seen as limited in different situations, e.g., while tele-operating a robot visual feedback may be directly provided by robot motion, and therefore a GUI may not be strictly necessary. 
Moreover, the concept of GUI is evolving because of the introduction of new visualisation techniques, for instance, those related to virtual or augmented reality, which as a matter of fact can be considered a whole separate field of research. 
For these reasons, and to devote much attention to a general analysis of gesture-based interaction, we decide to consider out of scope for this paper how system feedback is conveyed to the user. 

Interfaces mediated by tools such as keyboards or teach pendants, joysticks, and switches meant at controlling industrial robots are characterised by: (i) the operations defined by the class the tool belongs to, (ii) their specific layout, (iii) the user experience layer implementing the interface logic and (iv) a context-based feedback.
Similarly, the operation of a functional gesture-based interface is defined by different albeit correlated components, either physical or disembodied, which define how a gesture is perceived, which gestures the system is expected to react to, and how. 
The first step in the analysis of gesture-based interaction is obviously the definition of what a gesture is, how it can be characterised, and how such characterisation affects the structure of a gesture-based interface.

\subsection{Definition of Functional Gestures}
\label{sec:gesture_definition}

In general terms, it is not possible to provide an overall definition of what a gesture is, but it is possible to define it in the narrow scenario of HMI \cite{Pavlovic1997}. 
The notion of ``functional gesture'' results intuitive and, probably for this reason, in the literature the majority of works aiming at developing techniques and conceptual frameworks for gesture-based HMI do not explicitly define it. 
However, from an analysis of current state-of-the-art literature, we can extrapolate that, in HMI scenarios, functional gestures have been defined as \textit{trajectories} \cite{Pavlovic1997, Klboz2015} of \textit{body motion} \cite{Mitra2007, Tang2019, Pomboza-Junez2019} or \textit{poses} \cite{Chakraborty2018} \textit{performed intentionally} \cite{Kang2004, Ruffieux2015} with the intent of \textit{conveying meaningful information} or \textit{interacting with the environment} \cite{Mitra2007, Chakraborty2018, Tang2019, Pomboza-Junez2019}.
In the following paragraphs, we refer to functional gestures simply as gestures. 
Therefore, we can summarise state-of-the-art definitions by giving ours:

\begin{displayquote}
\textit{
Gestures are body actions that humans intentionally perform to affect the behaviour of an intelligent system.
}
\end{displayquote}

\noindent In providing this definition we have tried to be the more general as possible, with the aim of including all the different aspects which are present directly or indirectly in the literature\footnote{The focus of the definition is the action performed by a human and not the effects that it could have on other sensory media such as, for example, sound waves. 
Therefore, vocal utterances are not gestures.}. 
The usage of ``body actions'', substituting motions or poses, is meant at yielding a general definition without focusing on a particular kind of gesture.
Similarly, references to a particular body part, although in the literature gestures are usually defined for upper limbs, have been neglected in favour of a broader definition. 
Furthermore, the usage of the word ``intentionally'' is due to the functional aim of our definition. 
In fact, functional HMI implies that users are aware that the system they are interacting with is monitoring their actions, and a gesture, to be meaningful for the interaction, should be performed intentionally. 
Therefore, this aspect should be included in the definition since the machine is expected to be able to distinguish between gestures and unintentional actions. 
The generic reference to the ``behaviour'' of an intelligent system allows us to consider a definition encompassing different applications and different mapping about how a gesture could affect the system.

\begin{figure}[t!] 
    \centering    
    \includegraphics[width=.4\textwidth]{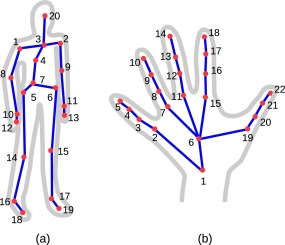}
    \caption[Full body and hand skeletons.]{Full-body (a) and hand (b) skeletons \cite{Nunez2018}.}
    \label{fig:skeleton}
\end{figure}

Inspired by those references whereby gestures are defined as trajectories in space, we extrapolated an analytical definition. 
If we represent the human body using a skeleton model such as the one in Figure \ref{fig:skeleton}, we can define human body postures using a joint status vector:
\begin{gather*}
\bm{q}(t) = \left\{q_1 \dots q_i \dots q_n \right\}
\end{gather*}

\noindent where $q_i$ is one of the general joint angles between two consecutive skeleton links (in human kinematics each joint can be characterised by multiple angles, e.g., the spherical joint of the shoulder), and $n$ is the number of the considered joint angles in the skeleton\footnote{it is worth noting that should it be necessary to specify the absolute pose of the human in space it would be possible to add virtual joints.}. 
Furthermore, we can define $\bm{\dot{q}}(t)$ and $\bm{\ddot{q}}(t)$ as the angular velocities and accelerations of each joint, respectively. 
Then, we can define a trajectory in a $n$-dimensional space containing joint angles, their angular velocities, and accelerations as:
\begin{gather*}
\bm{\tau}(t_s,t_e) = \left\{\bm{q}(t_s) \ldots	 \bm{q}(t_e), \bm{\dot{q}}(t_s) \ldots \bm{\dot{q}}(t_e), \bm{\ddot{q}}(t_s) \ldots	 \bm{\ddot{q}}(t_e) \right\}  
\end{gather*}

\noindent where $t_s$ is the time instant in which the trajectory starts and $t_e$ is when it ends. 
Therefore:
\begin{displayquote}
\textit{
Gestures are trajectories $\bm{\tau}(t_s,t_e)$ that humans intentionally perform to affect the behaviour of an intelligent system.
} 
\end{displayquote}

\noindent The substitution of the notion of ``action'' with that of a ``trajectory'' does not alter the generality of the definition as posed above.
In fact, as we will see in the next Section, while trajectories represent a motion naturally, they can be used to describe poses as well. 
It is worth noting that our definition is articulated in two components, namely the trajectory, and the behaviour of the intelligent system. 
This conceptual separation will prove to be particularly useful in the next Section when we will define gesture taxonomies.
 
\section{A Gesture Taxonomy}
\label{sec:2:taxonomy}

\subsection{Analysis of Existing Taxonomies}

Being able to establish appropriate gesture taxonomies is fundamental to properly define the kind of gestures a given HMI system is employing, and to have a clear understanding of the issues that such interface should address. 
Although in the literature this information is usually overlooked, characterising the kind of considered gestures is fundamental because it influences the synthesis of a well-defined problem statement for gesture-based interaction with machines.  
Before designing functional gesture-based interfaces, it is necessary to envisage a new taxonomy or select an existing one. 
Gestures are complex elements of an interface and can be described from many perspectives and considering different characteristics. 
In this Section, we analyse five characterising features discussed in the literature to propose conceptualisations of gesture taxonomies, namely \textit{time}, \textit{context}, \textit{level of instruction}, \textit{body part}, and \textit{spatial influence}.

The first gesture characteristic we consider is related to its temporal duration. 
From the temporal perspective, it is almost universally accepted that gestures can be classified as \textit{static} or \textit{dynamic} \cite{ChaoHu2004, Karam2005, Mitra2007, stern2008, Itkarkar2016, Jiang2017, Carreira2017, Buddhikot2018, Pomboza-Junez2019}.
Definitions can vary, but as a general rule-of-thumb static gestures are characterised by \textit{poses}, while dynamic ones by \textit{motions}. 
In the literature, dynamic gestures have been described as composed of three phases, namely \textit{pre-stroke} (or preparation), \textit{stroke}, and \textit{post-stroke} (or retraction) \cite{Quek1995, Mitra2007}. 
In the pre-stroke phase, a human moves to the starting pose associated with the gesture, during stroke the salient movement of the gesture happens, while in post-stroke either a resting position or the starting pose of the next gesture is reached.
Two additional phases can be added \cite{mcneill1992hand}, namely \textit{pre-stroke hold} (if stroke is delayed), and \textit{post-stroke hold} (if post-stroke is delayed).

The second gesture characteristic we discuss is related to the context associated with a gesture execution.
Context-oriented taxonomies generated in semiotic studies have been adopted in HMI, and from time to time adapted to specific scenarios. 
However, the lack of a shared taxonomy leads to a heterogeneous and quite diversified literature, where the same concept, notion, or term can have a varied meaning.
Nevertheless, a few similarities can be observed, which can contribute to a principled definition of a gesture. 
On a general basis, the dichotomy between gestures meant at communication (i.e., \textit{communicative} gestures), and those targeted at manipulation (i.e., \textit{manipulative} gestures) is widely accepted \cite{Quek1995, Pavlovic1997, Quek2002, Taralle2015, Vuletic2019}. 
As an example, Quek \textit{et al.} \cite{Quek1995} provide a compelling example of the differences between them: 

\begin{displayquote}
\textit{
An orchestral conductor’s hand motions are intended to communicate temporal, affective and interpretive information to the orchestra. 
A pianist’s hand movements are meant to perturb the ivories. 
While it may be possible to observe the pianist’s hands, the hand and finger movements are not meant to communicate with anyone.
}
\end{displayquote}

\noindent A recent, visual depiction on the communicative power of hand and arm gestures postulated by Quek \textit{et al.} \cite{Quek1995} as implemented in robots can be found in Alter 3, a humanoid torso from Ishiguro Labs characterised by human-like appearances acting as an orchestral conductor. 
The literature presents extensive classifications for communicative gestures, but only to a lesser extent for manipulative ones \cite{Quek1995, stern2008, Rojas-Munoz2019, Vuletic2019}.

Communicative gestures are further classified in \cite{Quek1995} as \textit{symbols} and \textit{acts}.
Symbols are gestures serving a linguistic role, and therefore requiring a common (typically mediated also by culture) background shared between the \textit{agent} performing the gesture and the one having to interpret it, being the latter a human or a machine. 
In this context, a symbol can be \textit{referential} or \textit{modalising}:
\begin{itemize}
\item referential gestures are irrespective of other (active or inactive) communication channels, and have a direct mapping into language significants or even words, e.g., rubbing the index and the thumb may refer to money in certain cultures; 
\item modalising gestures complement other active communication channels to add further information, e.g., holding hands apart while talking about an object may imply that it is long or short.
\end{itemize}

\noindent Acts are gestures whereby the associated motion is directly connected with the intended meaning.
An act can be classified as being \textit{mimetic} or \textit{deictic}:
\begin{itemize}
\item mimetic gestures represent common sense or familiar concepts usually performing a pantomime, e.g., mimicking the lighting up of a cigarette to ask for a lighter;
\item deictic gestures, also known as pointing gestures, can be further classified as \textit{specific} (if used to select a given object), \textit{generic} (if used to identify a class of objects by pointing to one of them), and \textit{metonymic} (when pointing to an object to refer to some entity related to it or its function).
\end{itemize}

Stern \textit{et al.} \cite{stern2008} extend the previous classification of manipulative and communicative gestures with other two kinds of gestures, namely \textit{control} and \textit{conversational} gestures:
\begin{itemize}
\item control gestures are used to command real or virtual objects, e.g., pointing gestures used to command a robot to pick up an object;
\item conversational gestures occur during verbal interaction and refer to speech-related content.
\end{itemize}

Rojas-Mu\~{n}oz and Wachs \cite{Rojas-Munoz2019} introduce the so-called MAGIC gestural taxonomy, and divide gestures in \textit{communicative}, \textit{manipulative}, \textit{Butterworth's}, and \textit{regulatory}.
While communicative and manipulative gestures have meanings similar to the descriptions given above, Butterworth's gestures are meant at signalling a failure in speech, e.g., the gesticulation of someone trying to recall or articulate a word, whereas regulatory gestures help control and better understand a conversation, e.g., understanding whose turn is to speak. 
It is noteworthy that the MAGIC taxonomy describes also nested classifications for communicative and regulator gestures.
However, since no clear description is provided for subclasses, we do not detail the analysis here.

Karam and Schraefel \cite{Karam2005} propose a slightly different taxonomy, dividing gestures in deictic, manipulative, \textit{semaphoric}, \textit{gesticulative}, \textit{linguistic}, and with \textit{multiple styles}:
\begin{itemize}
\item as also discussed above, deictic gestures are related to pointing movements used to establish the identity or the spatial position of an object;
\item manipulative gestures are meant at controlling any object by mapping the gesture movement to the location and pose of the object;
\item semaphoric gestures require a stylised dictionary of static or dynamic gestures usually associated with a command;
\item gesticulation motions are hand movements performed while a person is speaking;
\item linguistic gestures are meant at composing sentences in a language, e.g., in case of sign language for deaf people;
\item multiple styles are gestures without a specific focus and are composed of a variety of other kinds of gestures.
\end{itemize}

Finally, Vuletic \textit{et al.} \cite{Vuletic2019} try to uniform taxonomies already present in the literature. 
Their proposed taxonomy recognises the difference between manipulative and communicative gestures.
However, they divide communicative gestures in \textit{independent} and \textit{speech-related} ones.
Whilst independent gestures include symbolic, semaphoric, and pantomimic gestures, as described above, speech-related gestures complement speech contents, and are divided into \textit{iconic} (further divided into \textit{pictographic}, \textit{spatiographic}, \textit{kinematographic}, and \textit{metaphoric}), modalising, \textit{cohesive}, Butterworth's, \textit{adaptive}, and deictic.
Modalising, Butterworth's, and deictic gestures have been described above, whereas:
\begin{itemize}
\item iconic gestures complement information conveyed by speech to better illustrate what has been said, an example being a hand rolling motion while describing a rolling stone down a hill, and are further divided in \textit{pictographic}, i.e., describing shapes, \textit{spatiographic}, i.e., describing spatial relationships, \textit{kinematographic}, i.e., describing actions, and \textit{metaphoric}, i.e., for such abstract concepts as a cutting-like gesture used to interrupt a conversation;
\item cohesive gestures are performed to refer to concepts previously introduced in the conversation;
\item adaptive gestures are involuntary motions performed to release body tension.
\end{itemize}

The third gesture characteristic we focus on, i.e., the level of instruction, has been introduced in the review by Vuletic \textit{et al.} \cite{Vuletic2019}. 
From the perspective put forth in that review, gesture-based HMI can be classified in \textit{prescribes}, i.e., those interfaces in which a gesture dictionary is required, and \textit{free-form}, i.e., a dictionary is not necessary.
It is noteworthy that here a focus shift is present.
In fact, all previous taxonomies are gesture-centred, while when considering the level of instruction and the taxonomy proposed in the review, what is classified is the whole interaction.


In our opinion, the features characterising gestures in HMI interfaces influence the problem statement, but not all differences in gesture features imply a difference in the problem statement itself.
Just for the sake of the argument, a machine used to interpret sign languages or semaphoric gestures should face the same engineering issues: (i) continuously sensing human actions, (ii) isolating relevant part of the data stream to prevent unintentional movements to be recognised as functional gestures, (iii) classifying the selected data according to a dictionary, and (iv) associating a meaning to the performed sign gesture, or a command for semaphoric gestures.
On the contrary, if we consider a manipulative gesture, human actions should be continuously sensed, the data stream be processed to extract relevant features, e.g., the hand orientation, and those features be continuously mapped to a relevant machine function, e.g., the orientation of an object in a 3D virtual environment.
These are two cases whereby the selected gestures affect data processing, but other differences can affect other aspects of the problem statement. 
For example, a gesture performed with an arm can be sensed using a wrist-worn Inertial Measurement Unit (IMU) but in order to sense a gesture performed with the fingers, another device should be used. 

The latter example allows us to introduce the fourth gesture characteristic we consider, i.e., the body part used to perform a gesture. 
This characteristic is usually ignored in the literature since gestures, especially in HMI interfaces, are implicitly considered to be performed with upper limbs. 
For this reason, apart from very few cases, in the literature, the specification of the used body part is completely neglected, which can be a major cause of confusion. 
Although it is implicit that gestures are performed with upper limbs, it is not always clear if a specific gesture requires using an arm, a hand, or the fingers. 
Only a few works clearly specify their focus on hand-based gestures \cite{Kela2006} or organise gestures as \textit{hand-arm}, \textit{head-face} and \textit{body} \cite{Mitra2007}.

Finally, the last gesture characteristic described and used to classify gesture-based interaction is spatial influence \cite{Mitra2007}. 
According to this perspective, the same gesture may have a different meaning depending on where it is performed.
For example, pressing a mid-air virtual button it is always performed in the same way, however, the position of the hand determines which button is actually pressed \cite{Shin2017non-touch}.

\subsection{A Reasoned Taxonomy of Gestures}

In this Section, building on the knowledge about available, state-of-the-art, gesture taxonomies introduced above, we can define and describe the components of the taxonomy we introduce in this paper. 
As a preliminary note, we stress out that our taxonomy is focused on the issues that each gesture characterisation implies, where no sociological connotation is involved. 
Furthermore, the taxonomy does not present a hierarchical structure.
In the discussion that follows, we first define the main characteristics, and then we analyse how they affect the problem statement.
We consider four features, which we refer to as \textit{effect}, \textit{time}, \textit{focus}, and \textit{space}\footnote{\url{https://youtu.be/-IQ4lOZ-BNI}}.

The effect is aimed at describing how a gesture is going to affect the machine the human is interacting with. 
If we refer to the function mapping a gesture $\bm{\tau}$ to the system state as $f(\bm{\tau})$, its effect describes the mapping itself.
According to their effect, gestures can be either continuous or discrete.
The information yield by a continuous gesture is mapped at each time instant to a change in the interface state, while for discrete gestures such change is \textit{atomically} associated with the whole gesture.
A gesture effect has also implications on the system state, which should be continuous or discrete according to the selected kind of gestures. 
Let us define $S_c \in \mathbb{R}^n$ as an $n$-dimensional continuous state, and $S_d \in \mathbb{S}$ as a discrete state, where $\mathbb{S}$ the set of all possible, countable, discrete states. 
For continuous gestures, $f(\bm{\tau})$ has domain $\bm{\tau}^{(t_e - t_s)}$ and co-domain $S_c^{(t_e - t_s)}$, whereas for discrete gestures the co-domain is simply the discrete state $S_d$. 
The two time instances $t_s$ and $t_e$, defined in Section \ref{sec:gesture_definition}, refer to the start and end events of a gesture.
Obviously, and adopting a pragmatic perspective, $S_c$ can not be ideally \textit{continuous} since its variations are related to the frequency at which the gesture is sampled by the employed sensors. 
If we consider a parallelism with the mouse-keyboard interface, continuous gestures are mouse movements while discrete gestures are the keystrokes.
It is noteworthy that although the classification between continuous and discrete gestures could recall the dichotomy between manipulative and communicative gestures, they are intrinsically different.
The context characteristic described above has been used to describe the intended meaning of gestures, for example, if they are aimed at manipulation or communication, while here we are describing the nature of the gesture effect. 
As an example, let us consider two different communicative gestures, i.e., semaphoric and language gestures. 
The former class includes gestures whose effect is defined on the basis of a synthetic dictionary associated with the execution of a command, while the latter is related to gestures characterised by a linguistic role. 
This difference in their usage characterises their distinction from a contextual perspective.
However, from an effect-related perspective, as long as these gestures are associated with a discrete command and a word, respectively, they are both discrete gestures.


In continuity with state-of-the-art taxonomies, the time characteristic classifies gestures as static or dynamic. 
However, differently from what is typically postulated in the literature, where phases are only foreseen for dynamic gestures, we organise all gestures in three phases, i.e., \textit{preparation}, \textit{stroke} and \textit{retraction}. 
A classification of a gesture as static or dynamic is aimed at describing the human behaviour during the stroke phase.
Therefore, in our case, a person performing a static gesture first moves to the desired pose (preparation), then keeps a static pose for an arbitrary amount of time (stroke), and finally returns to the rest position or reaches the starting pose of the next gesture (retraction). 
If we refer to the gesture definition introduced in Section \ref{sec:gesture_definition}, we can specify that 
for dynamic gestures it exists at least one $q_i \in \bm{q}$ such that $\dot{q_i} \neq 0$, whereas for static gestures it holds that $\bm{\dot{q}}(t_s) = \bm{\ddot{q}}(t_s) = 0$. 
Please notice that $t_s$ and $t_e$ refer here to the starting and ending time of the stroke phase.

Our taxonomy enforces that static gestures cannot be considered continuous.
In fact, although a static gesture \textit{per se} may be considered continuous, it is not possible to associate a static gesture to a continuous system state.
To make this possible, it would be necessary to define an infinite amount of postures for each possible instance element of the continuous system state, and this is not only impractical but \textit{de facto} impossible.

In the taxonomy, focus describes which body parts are relevant for a gesture. 
The name of this characteristic has been chosen to highlight the fact that the relevance of a body part is determined \textit{a priori} by the HMI requirements, and it does not depend on the specific action. 
For example, in an application whereby the focus is on the arm, whichever gesture is performed by the hand is ignored. 
The focus characteristic does not divide gestures in distinct classes, but describes each gesture using the names of the relevant body parts. 
Referring again to our gesture definition, the focus on a specific body part implies that we are interested only in the status of the joints relevant to that body part, e.g., in a hand gesture, the interested joints are the ones modelling the wrist, essentially. 
For this reason, a waving gesture is first of all an arm gesture, since the motion is generated by arm joints, and can be considered a combined arm-hand gesture if we are interested even in the status of the wrist.

Considering only relevant body parts helps reduce the size of an associated classification problem.
In fact, if just a subset $\bm{p}(t) \subseteq \bm{q}(t)$ is needed to represent a gesture, then the trajectory associated with that gesture can be redefined as $\bm{\tau}_p(t_s, t_e)$. 
It is noteworthy that focus can also model gestures referring to multiple, or even non directly connected, body parts, e.g., bi-manual gestures \cite{shah2018detection}. 

The last gesture characteristic we include, namely space, likewise other taxonomies described in the previous Section, determines whether the meaning associated with a gesture is influenced by the physical location of the body part performing it. 
According to this distinction, gestures can be spatially related or unrelated. 
We have previously seen how the discrete gesture of a mid-air virtual button pressure can be spatially related. 
Similarly, a manipulative gesture for dragging and dropping a virtual object is spatially related, however a manipulative gesture for controlling the velocity of a mobile robot by means of the arm inclination and rotation, as done in \cite{Coronado2017}, is spatially unrelated.

The four characteristics described in the Section make up our gesture taxonomy.
A human action can be composed of multiple gestures. 
Getting back to the hand waving example, if we want to describe the fact that the fingers are spread while the arm is moving, then the gesture can be classified as a \textit{discrete gesture dynamic-arm and static-fingers spatially unrelated}, otherwise it could be simply a \textit{discrete gesture dynamic-arm spatially unrelated}. 

\section{Problem Statement}
\label{sec:problem_statement}

In the previous Section, we have introduced our notion of (functional) gesture and described its characterisation in the form of a taxonomy. 
Here, we aim at structuring the problem of gesture-based interfaces for human-machine interaction. 
As anticipated in the Introduction, we refer to this problem as Gestural Interaction for User Interfaces, and we refer to it as GI-UI.
It is noteworthy that the problem statement we put forth in this paper has an operational and engineering nature.
As such, it is structured as the interplay among three, interrelated, sub-problems, i.e., \textit{sensing}, \textit{data processing} and \textit{system reaction}. 
Although the structure of the problem statement is fixed and well-defined, i.e., sensing influences data processing, which is mapped to a certain system reaction, the design choices of sub-problems are tightly related to the gestures an HMI interface can consider.
In the following paragraphs, we explore the relationships among these sub-problems more in-depth.

\begin{table}[t!]
\caption[Gestures' influence on problem components.]{Summary of the influence that gestures can have on sensing, data processing, and system reaction according to gesture type.}
\centering
\label{table:probletaxonomyrelation}
\begin{tabular}{c|c|c|c}
\textit{Characteristic} & \textit{Sensing}  & \textit{Data Processing}  & \textit{System Reaction}  \\ \hline \rule{0pt}{2.6ex}
Effect                  & None              & Major                     & Major                     \\ 
Time                    & Major             & Minor                     & None                      \\ 
Focus                   & Minor             & Minor                     & None                      \\ 
Space                   & Major             & Minor                     & None                
\end{tabular}
\end{table}

\subsection{Sensing}
\label{sec:sensing}

The kind of sensors used to collect data about gestures depend on the particular application requirements, e.g., privacy \cite{zhao2015privacy}, price \cite{Wachs2011} or computational efficiency \cite{poularakis2015low}.
Two conflicting paradigms affect gesture sensing, namely \textit{come as you are} and \textit{wearability} \cite{Wachs2011}. 
The phrasing ``come as you are'' refers to a system design whereby its users should not \textit{wear} anything to interact with the machine favouring a more natural interaction.
On the contrary, wearability refers to systems assuming that users are wearing such interaction tools as data gloves or smartwatches, i.e., they bring the HMI interface with them. 
These two paradigms have specific advantages and drawbacks and should be selected according to the application.
Furthermore, gesture sensing approaches are typically classified as non-image-based and image-based \cite{Liu2018}. 
Non-image-based methods include both wearable devices, such as instrumented gloves, wristbands and body suits, and non-wearable devices, using radio frequency or electric field sensing. 
Image-based approaches are probably one of the most widely explored research fields, as many literature reviews address specific issues and solutions related to this sensing modality \cite{Itkarkar2016, Cheng2016}.
However, the list of sensing solutions \cite{Vuletic2019} is wide and not easy to analyse nor classify. 
Our aim is not to list different approaches or to propose a classification. 
Instead, what we want to highlight is the influence that the kind of gestures we consider can have in the choice of the sensing device, while taking into consideration that -- as we have seen previously -- other factors may affect our choice as well. 
A visual representation of the extent to which gesture characteristics may influence the sensing strategy is presented in Table \ref{table:probletaxonomyrelation}. 
According to our observations, the temporal and spatial characteristics have a significant influence on the sensing strategy.
Although usually it is possible to sense static and dynamic gestures using the same sensors, some sensors are more suited for specific kinds of gestures.
For instance, in stationary conditions, an accelerometer can be used to determine its inclination with respect to the horizontal plane, therefore making it easy to monitor simple static gestures \cite{brodie2008static}. 
Instead, if we consider a dynamic gesture the usage of accelerometers is not enough to track the sensor pose and information from other sensors, such as gyroscopes or magnetometers, should be integrated \cite{tseng2011motion}. 
Similarly, in order to recognise a spatial gesture, such as a pointing gesture, one must select a sensor allowing for the extraction of spatial information, e.g., a camera \cite{Lai2016}. 
Finally, we can observe that the influence of focus on the sensing strategy is limited, and it is mainly associated with wearable devices. 
In fact, depending on the gesture focus, the sensor should be placed to have visibility of the movement, e.g., for a gesture involving fingers IMUs should be placed on the fingers and not on the arm \cite{Xie2015}.


\subsection{Data Processing}
\label{sec:data_processing}

Many factors can influence data processing.
One in particular, however, drastically changes the nature of the problem an HMI designer must solve, i.e., the effect, as presented in Table \ref{table:probletaxonomyrelation}. 
As a matter of fact, depending whether we consider continuous or discrete gestures, the problem statement completely changes. 
In the case of continuous gestures, at each instant its representation must be directly associated with a machine reaction, or function. 
In certain cases, this is possible using raw data \cite{Coronado2017}. 
However, data should be processed to extract relevant, possibly \textit{semantic} features, such as the 2D position of the hand with respect to an image plane \cite{Bolano2018}. 
As it is customary, we refer to this procedure as \textit{feature extraction}.
Instead, for discrete gestures feature extraction is part of a more complex problem. 
Raw data or the extracted features should be analysed to determine the start and the end points of the gesture, and to classify it according to a predefined dictionary \cite{Carfi2018}, a process usually termed \textit{gesture recognition}. 

Gesture recognition has been described as the process whereby specific gestures are recognised and interpreted \cite{Mitra2007, Ruffieux2015}.
Some studies consider gesture recognition as a three-step process, composed of identification, tracking and classification \cite{Liu2018}, or alternatively of detection, feature extraction and classification \cite{Chakraborty2018}.
Other studies consider it a two-step process, made up of detection and classification \cite{Naguri2017}. 
In the literature, the characterisation of the gesture recognition problem is highly influenced by the considered sensors and the target application. 
Here, we try to give a general definition. 
The gesture recognition problem is the process that, given sensory data and a dictionary of discrete gestures, determines whether any gesture has occurred, and which one.
To this aim, sensory data are supposed to undergo three computational steps, namely pre-processing and feature extraction, detection, and classification. 
In the first phase, raw sensory data are manipulated to de-noise and to extract relevant features.
In the detection phase (also referred to as segmentation or spotting), filtered data are analysed to determine the occurrence of a gesture, and its start and end moments.
Detection is usually performed using filtering techniques or threshold-based mechanisms \cite{Liu2018}. 
The classification phase determines which gesture present in the dictionary has been performed.  
Often, classification is probabilistic, and together with the label it returns a confidence value \cite{Carfi2018}. 
The literature has explored different approaches for gesture recognition, encompassing purely model-based techniques \cite{Kortenkamp1996} to machine learning \cite{Islam2019}, whereby the adopted techniques are highly dependent on the data source. 
The order used to discuss the three phases is not binding, especially with reference to detection and classification. 
In fact, gesture detection can be direct, when it is performed on sensory data, or indirect, when it is performed on classification results \cite{Escalera2017}. 

As shown in Table \ref{table:probletaxonomyrelation}, other gesture characteristics can have minor effects on data processing. 
Static and dynamic gestures imply intrinsically different data, since gestures belonging to the former class are not related to time, whereas dynamic gestures are.
Therefore, the techniques adopted to process static and dynamic gestures are different. 
Similarly, the focus of a gesture influences the kind of collected data, or the features that the system should extract, i.e., \textit{head} or hand gestures extracted from video streams imply a different data processing pipeline.
Finally, spatially related gestures require the feature extraction process to consider the position of the body part performing the gesture as a feature. 


\subsection{System Reaction}
\label{sec:system_reaction}

The way an HMI system responds and acts to a given gesture varies depending on the application.
As an example, it may consist of a switch in an interface menu (HCI) \cite{Gupta2012}, or a velocity command for a mobile robot (HRI) \cite{Coronado2017}.
Obviously enough, system responses are the results of a mapping between data processing and machine behaviour.
This mapping serves as a sort of semantic attribution to gestures in the context of the interaction process.
As it can be seen in Table \ref{table:probletaxonomyrelation}, the reaction is to a large extent agnostic to gesture characteristics, since their effects are absorbed by the sensing and the data processing phases. 
The effect characteristic affects of course system reaction, since continuous and discrete gestures by their own nature implies continuous and discrete system functions, respectively.

An obvious topic to deal with when considering machine behaviours in response to human gestures in HMI interfaces is how the machine can expose a natural and intuitive interface to its users.
Many studies have been carried out in the past decades, and the recent urge to design and develop technologies for the consumer market has further increased this trend \cite{Vuletic2019}.
Therefore, we decided not to consider this topic, and we deem it out of scope. 
The interested reader is referred to \cite{Wachs2011}.
However, we focus in the next Section on those interaction-related traits that can be directly linked to the taxonomy we introduce in Section \ref{sec:2:taxonomy}, and to the problem statement presented in \ref{sec:problem_statement}.

\section{Human Factors in Gesture-based Interfaces}
\label{sec:humanfactor}

\subsection{Requirements}
\label{sec:requirements}

In the previous Sections, we have considered humans as mere entities performing gestures, and we have discussed HMI processes only from a machine perspective.
However, human presence is fundamental and can influence the design choices taken while developing solutions for the GI-UI problem as a whole, although in particular system reaction is affected. 
Many requirements dealing with human factors, and well-aligned with the GI-UI problem, have been identified in the literature \cite{Wachs2011}.
These include \textit{responsiveness}, \textit{user adaptability and feedback}, \textit{learnability}, \textit{accuracy}, \textit{low mental load}, \textit{intuitiveness}, \textit{comfort} and \textit{lexicon size}. 
All these requirements should be taken into account while designing an HMI process.
Although they do not modify the general structure of the problem, they can surely enforce certain solutions with respect to others. 

Responsiveness is a metric typically associated with the dynamic flow of the HMI interface.
As a rule-of-thumb, the \textit{response time}, i.e., the time interval between user input and system reaction, should be as low as possible. 
The overall system should be carefully designed to react to gestural stimuli as fast as possible \cite{ChaoHu2004, Lee2014}.
Even a non expert user can easily perceive an increased reaction time, and if it exceeds a psychological and cognitive threshold, user experience and satisfaction can be seriously disrupted \cite{Attig2017}. 
Different applications may imply different such thresholds. 
For general-purpose applications, classical design methodologies in HMI recommend response times lower than $100$ ms.
However, recent studies found out that the acceptable latency may be even lower, especially for interactions resembling physical ones \cite{Attig2017}. 
This requirement may prevent designers to adopt sensors generating lots of data, or particularly heavy data, especially when the processing hardware is not top performance, such as in edge or field applications or when the associated processing techniques are characterised by a high level of complexity.

Especially in HMI interfaces employing discrete gestures, the system is supposed to distinguish among a limited number of gestures. 
Therefore, depending on the application, the interface should be capable of \textit{adapting} to certain user specific traits, either cognitive or physical \cite{carreira2017evaluation}.
Many contributions to the literature highlight the need for gesture-based interfaces to allow a user to personalise the (set of) gestures used in the interaction \cite{Kela2006}. 
This requirement has an obvious technological consequence, i.e., the techniques used to model gestures, as well as those related to its run-time processing, must allow for an easy-to-attain adaptation even for non expert users. 
Most likely, this would involve 
the possibility for the system to learn from experience \cite{Kuno2000}.

Furthermore, the HMI interface is expected not only to react to the detected gesture minimising the response time, but also to provide the user with an adequate \textit{feedback} about the correctness of the gesture itself.
Lack of direct feedback in HMI can ultimately lead to a lack of trust in the system and a deranged user experience \cite{Wachs2011}.
Among the methods currently exploited to provide feedback to users, we can mention classical GUIs \cite{Bolano2018}, haptics \cite{kim2015depth}, augmented \cite{bai2013free} and virtual \cite{Park2019} reality.

For a real-world use, a gesture-based interface system correctly interprets gestures in input with \textit{accuracy} levels close to $100$\% \cite{Kela2006}. 
This is not only preferred from an engineering perspective, but also of the utmost importance for an interface aimed at being used in everyday conditions. 
As a direct, although not obvious consequence of this requirement, gestures should be designed and modelled with two somewhat contrasting objectives in mind. 
On the one hand, they should be such to maximise the ease of classification, i.e., enforcing accuracy; on the other hand, they should preserve their intuitiveness \cite{Islam2019}. 
In accordance with our definition of continuous and discrete gestures, the notion of accuracy for these two kinds of gestures is different. 
In fact, while for continuous gestures it is important to properly estimate the relevant features, and therefore depending by the specific feature the error metric can change, in the case of discrete gestures what matters most is the recognition rate, usually expressed using such parameters as accuracy, precision, recall, and the F1 score \cite{ward2011performance}.

The requirements associated with learnability, mental load, comfort and intuitiveness are strictly intertwined.
The set of gestures should be easy to learn, whereas the training time for new users should be brief \cite{Mantyjarvi2004}. 
While in a general sense the HMI process, and the interface in particular, should not heavily impact on the user mental load \cite{Wachs2011}, it is widely accepted that gesture-based HMI can reduce it to a great extent \cite{Taralle2015}.
As a consequence, the selected gestures should be simple and brief, which is expected to enforce also learnability. 
Complex gestures, or gestures that imply intense muscle activity should be avoided to guarantee user comfort, especially if the gesture-based interface is designed to be used for a prolonged time \cite{Wachs2011}
Finally, selected gestures should have an intuitive association with the expected system behaviour.
This, in turn, facilitates learnability and reduces the user mental load \cite{Gonzalez2018}.





The capability of a gesture-based HMI interface to interpret a high number of gestures, i.e., the lexicon size, can be of the utmost importance for its usability and effectiveness, an obvious example being the automated interpretation of the sign language \cite{Galka2016}.
However, increasing the lexicon size is in contrast with the learnability and low mental load requirements, because a bigger dictionary is more difficult to learn and recall \cite{Yeasin2000}. 
Furthermore, a bigger dictionary may imply lower performance in data processing, because of increased problem complexity. 
As a consequence, the size of the dictionary should be mediated by designers considering all these factors.

In summary, we can observe that many of the considered requirements refer to the selected gestures and to the implications that their choice can have both on user experience and the problem solution. 
We can conclude from this brief overview that the gesture choice and the associated system behaviour are fundamental. 
The lexicon size is specifically related to the dictionary composition.
As we have previously mentioned, dictionary design is a problem associated with gesture recognition and therefore to the specific case of discrete gestures. 
In the next Section we focus on what a gesture dictionary is, and how we can design it. 

\subsection{Gesture Dictionaries}
\label{sec:chapter2:dictionary}

In general terms, a dictionary (or vocabulary) is the set of words making up a language and the associated meaning.
In our case, the dictionary contains the set of gestures that the interface is able to interpret, i.e., for each gesture a description of the trajectory $\bm{\tau}$, as well as the associated elicited behaviour. 
The definition of a gesture dictionary is an issue related only to semaphoric gestures, as introduced in Section \ref{sec:2:taxonomy}. 
For all other gestures, a dictionary is still necessary, but as a matter of fact it is given by the context. 
For example, the development of an interface for the translation of sign language does not require the definition of a new dictionary since it is already available.

For semaphoric gestures, the dictionary can be defined as a set of matched pairs of commands and their gestural expression \cite{stern2008}. 
As we already discussed in the previous Section, a dictionary should made up of gestures that are intuitive, ergonomic, easy to be recognised, easy to learn, and easy to remember \cite{stern2008}. 
In order to build dictionaries, for a given application and to meet certain requirements, many alternatives have been explored in the literature. 
Different approaches can be used to build dictionaries.
The first approach implies using questionnaires whereby volunteers are asked to sketch the gestures they consider most appropriate for the specific application.
It has been observed that if the gestures are intuitive enough, a user can easily adapt to gestures defined by others \cite{Kela2006}. 
According to the second approach, volunteers can be required to mimic a gesture-based interaction for a specific application, and as a consequence the dictionary is built based on the experimenter observations \cite{Jahani2018}. 
The third approach involves the usage of Wizard of Oz experiments \cite{Hoysniemi2005}. 
In this case, volunteers are tasked with solving an interaction problem with a machine they suppose to be autonomous.
Instead, the machine is manually operated by the experimenter to mimic its behaviour in response to human gestures as if the machine was autonomous.
Then, the observations of gestures done by volunteers can be used to define the dictionary. 

Once the dictionary is designed, it is important to have tools to determine whether it satisfies the requirements listed above. 
This can be achieved by performing experiments and determining metrics to evaluate the considered characteristics \cite{stern2008}. 
Alternatively, a few studies have proposed the Vocabulary Acceptability Criteria (VAC) \cite{Gonzalez2018}, which allows experts to evaluate gestures on the basis of six attributes, namely \textit{iconicity}, i.e., how much a gesture recalls the associated command, \textit{simplicity}, \textit{efficiency}, \textit{compactness}, i.e., how much the gesture covers the space around the body, \textit{salience}, i.e., how discriminating a movement is, and \textit{economy}, i.e., related to the movement magnitude \cite{Gonzalez2018}. 
Although it is nowadays clear what are the requirements that should characterise a dictionary, no standard exists yet for dictionary design and evaluation. 

Another problem related to gesture dictionaries is how to effectively describe a gesture. This is important for reproducible research.
Often, original data sets are not available, and the only way to reproduce work done by others is to collect a new data set following the same experimental procedure. 
However, it is not always clear how gestures have been performed. 
Gesture taxonomies can help disambiguate in some scenarios, but they may not be sufficient. 
Drawings in the literature are often used to describe the gestures and are preferred to videos because they can be easily shared and printed.
We believe that drawing combined with a text-based description is the most informative solution as far as scholarly value is concerned.
However, this still remains an open issue.

\begin{table*}[!b]
    \caption[Gestural literature classification (table 1/2).]{Classification of literature related to gesture-based interfaces.}
    \centering
    \label{table:GestureClassification}
    \resizebox{\textwidth}{!}{\begin{tabular}{c c c c c c c c c c c c c}
    \multicolumn{2}{c}{\textit{Article info}} & \multicolumn{3}{c}{\textit{Sensing}} & \multicolumn{1}{c}{\textit{Reaction}} & \multicolumn{2}{c}{\textit{Processing}} & \multicolumn{5}{c}{\textit{Gestures}}\\
    \textit{Ref} & \textit{Year} &  \textit{Sensor (1)} & \textit{Sensor (2)} & \textit{Sensor (3)} &  & \textit{Problem} & \textit{User defined} & \textit{Size} & \textit{Effect} & \textit{Time} & \textit{Focus} & \textit{Space}\\  
    \toprule


    \multirow{3}{*}{\cite{Kortenkamp1996}} & \multirow{3}{*}{1996} & Stereo & \multirow{3}{*}{-} & \multirow{3}{*}{-} & \multirow{3}{*}{HRI} & \multirow{3}{*}{Gesture Recognition} & \multirow{3}{*}{No} & \multirow{3}{*}{6} & \multirow{3}{*}{Discrete} & \multirow{3}{*}{Static} & \multirow{3}{*}{Arm} & \multirow{3}{*}{Yes} \\
    & & Monochrome \\
    & & Camera \\

    \midrule
    \multirow{3}{*}{\cite{Cipolla1996}} & \multirow{3}{*}{1996} & Stereo & \multirow{3}{*}{-} & \multirow{3}{*}{-} & \multirow{3}{*}{HRI} & \multirow{3}{*}{Feature Extraction} & \multirow{3}{*}{No} & \multirow{3}{*}{-} & \multirow{3}{*}{Continuous} & \multirow{3}{*}{Dynamic} & \multirow{3}{*}{Fingers} & \multirow{3}{*}{Yes} \\
    & & Monochrome \\
    & & Camera \\

    \midrule
    \multirow{2}{*}{\cite{Starner2000}} & \multirow{2}{*}{2000} & Monochrome & Infrared & \multirow{2}{*}{Buttom} & \multirow{2}{*}{-} & Gesture Recognition & No & 8 & Discrete Continuous & Static Dynamic & Fingers Arm & No\\
    & & Camera & Illumination & & & Gesture Recognition & Yes & 6 & Discrete & Dynamic & (Bi) Fingers Hand & No\\

    \midrule
    \cite{Yeasin2000} & 2000 & Camera & - & - & - & Gesture Recognition & No & 5 & Discrete & Dynamic & Finger Hand & No\\

    \midrule
    \cite{Kuno2000} & 2000 & Camera & - & - & HRI & Gesture Recognition & Yes & - & Discrete & Dynamic & (Bi) Hand & No\\ 
    
    \midrule
    \multirow{2}{*}{\cite{su2000}} & \multirow{2}{*}{2000} & Electro-mechanical & \multirow{2}{*}{-} & \multirow{2}{*}{-} & \multirow{2}{*}{-} & \multirow{2}{*}{Gesture Classification} & \multirow{2}{*}{No} & \multirow{2}{*}{90} & \multirow{2}{*}{Discrete} & \multirow{2}{*}{Dynamic} & \multirow{2}{*}{(Bi) Fingers} & \multirow{2}{*}{No} \\
    & & Strain Gauges \\

    \midrule
    \cite{Rogalla2002} & 2002 & Camera & - & - & - & Gesture Classification & No & 6 & Discrete & Static & Fingers Hand & No \\ 

    \midrule
    \cite{Ramamoorthy2003} & 2003 & Camera & - & - & - & Gesture Recognition & No & 5 & Discrete & Dynamic & Fingers Hand & No \\ 

    \midrule
    \cite{ChaoHu2004} & 2003 &  RGB Camera & - & - & - & Gesture Classification & No & 6 & Discrete & Static & Fingers Hand & No \\

    \midrule
    \multirow{2}{*}{\cite{Hasanuzzaman2004}} & \multirow{2}{*}{2004} & \multirow{2}{*}{RGB Camera} & \multirow{2}{*}{-} & \multirow{2}{*}{-} & \multirow{2}{*}{HRI} & Gesture Recognition & No & 8 & Discrete & Static & (Bi) Fingers Hand & No\\
    & & & & & & Gesture Recognition & No & 2 & Discrete & Dynamic & Head & No\\ 

    \midrule
    \cite{Mantyjarvi2004} & 2004 &  Accelerometer & Button & - & - & Gesture Recognition & No & 8 & Discrete & Dynamic & Hand & No \\

    \midrule
    \cite{Kang2004} & 2004 &  RGB Camera & - & - & HCI & Gesture Recognition & No & 10 & Discrete & Dynamic & (Bi) Arm Torso & No \\

    \midrule
    \multirow{2}{*}{\cite{Kela2006}} & \multirow{2}{*}{2006} &  \multirow{2}{*}{Accelerometer} & \multirow{2}{*}{Button} & \multirow{2}{*}{-} & \multirow{2}{*}{HCI} & Gesture Recognition & No & 8 & Discrete & Dynamic & Hand & No \\
    & & & & & & Feature Extraction & No & - & Continuous & Dynamic & Hand & No\\

    \midrule
    \multirow{2}{*}{\cite{Nickel2007}} & \multirow{2}{*}{2007} & Stereo & \multirow{2}{*}{-} & \multirow{2}{*}{-} & \multirow{2}{*}{-} & \multirow{2}{*}{Gesture Recognition} & \multirow{2}{*}{No} & \multirow{2}{*}{1} & \multirow{2}{*}{Discrete} & \multirow{2}{*}{Dynamic} & \multirow{2}{*}{Hand} & \multirow{2}{*}{Yes}\\
    & & Camera & & & & & & & & & & \\

    \midrule
    \multirow{2}{*}{\cite{Kim2007}} & \multirow{2}{*}{2007} & Proximity & \multirow{2}{*}{-} & \multirow{2}{*}{-} & \multirow{2}{*}{-} & \multirow{2}{*}{Gesture Recognition} & \multirow{2}{*}{No} & \multirow{2}{*}{5} & \multirow{2}{*}{Discrete} & \multirow{2}{*}{Dynamic Static} & \multirow{2}{*}{Hand} & \multirow{2}{*}{No}\\
    & & Sensors \\

    \midrule
    \cite{Bailador2007} & 2007 &  Accelerometer & Button & - & - & Gesture Recognition & No & 8 & Discrete & Dynamic & Hand & No \\



    \midrule
    \cite{Schlomer2008} & 2008 &  Accelerometer & Button & - & - & Gesture Recognition & No & 5 & Discrete & Dynamic & Hand & No \\

    \midrule
    \multirow{2}{*}{\cite{Neto2009}} & \multirow{2}{*}{2009} &  \multirow{2}{*}{Accelerometer} & \multirow{2}{*}{-} & \multirow{2}{*}{-} & \multirow{2}{*}{HRI} & Gesture Classification & No & 12 & Discrete & Dynamic & Arm & No \\
    & & &  & & & Gesture Recognition & No & 2 & Discrete & Static & Arm & No\\

    \midrule
    \multirow{2}{*}{\cite{Liu2009}} & \multirow{2}{*}{2009} & \multirow{2}{*}{Accelerometer} & \multirow{2}{*}{Button} & \multirow{2}{*}{-} & \multirow{2}{*}{HCI} & Gesture Recognition & No & 8 & Discrete & Dynamic & Hand & No\\
    & & &  & & & Gesture Recognition & Yes & - & Discrete & Dynamic & Hand & No\\

    \midrule
    \cite{Xing-HanWu2010} & 2010 &  Accelerometer & - & - & HRI & Gesture Recognition & No & 6 & Discrete & Dynamic & Arm & No \\

    \midrule
    \cite{Akl2010} & 2010 &  Accelerometer & Button & - & - & Gesture Recognition & No & 18 & Discrete & Dynamic & Hand & No \\
    
    \midrule
    \cite{ni2011} & 2011 &  MoCap & Button & - & HCI & Feature Extraction & No & - & Continuous & Dynamic & Arm & No \\
    
    \midrule
    \multirow{2}{*}{\cite{zhang2011}} & \multirow{2}{*}{2011} &  \multirow{2}{*}{Accelerometer} & \multirow{2}{*}{EMG} & \multirow{2}{*}{-} & \multirow{2}{*}{HCI} & Gesture Recognition & No & 72 & Discrete & Dynamic & Hand Fingers & No \\
     &  &  &  & &  & Gesture Recognition & No & 18 & Discrete & Dynamic Static & Hand Fingers & No \\
    
    \midrule
    \multirow{3}{*}{\cite{Chen2012}} & \multirow{3}{*}{2012} &  \multirow{3}{*}{Accelerometer} & \multirow{3}{*}{Button} & Marker & \multirow{3}{*}{-} & \multirow{3}{*}{Gesture Recognition} & \multirow{3}{*}{No} & \multirow{3}{*}{20} & \multirow{3}{*}{Discrete} & \multirow{3}{*}{Dynamic} & \multirow{3}{*}{Hand} & \multirow{3}{*}{No} \\
    & & & & Based \\
    & & & & MoCap\\

    \midrule
    \cite{Khan2012} & 2012 &  Accelerometer & - & - & - & Gesture Classification & No & 8 & Discrete & Dynamic & Hand & No \\

    \midrule
    \cite{Gupta2012} & 2012 &  Microphone &  Speaker & - & HCI & Gesture Recognition & No & 5 & Discrete & Dynamic & (Bi) Hand & No \\
    
    \midrule
    \multirow{2}{*}{\cite{ruppert2012}} & \multirow{2}{*}{2012} &  \multirow{2}{*}{RGB-D Camera} & \multirow{2}{*}{-} & \multirow{2}{*}{-} & \multirow{2}{*}{HCI} & Gesture Recognition & No & 4 & Discrete & Dynamic & Hand & No \\
     &  &  &  & &  & Feature Extraction & No & - & Continuous & Dynamic & Hand & Yes \\
    
    \midrule
    \multirow{2}{*}{\cite{Porzi2013}} & \multirow{2}{*}{2013} &  \multirow{2}{*}{Accelerometer} & Touch & \multirow{2}{*}{-} & \multirow{2}{*}{HCI} & \multirow{2}{*}{Gesture Recognition} & \multirow{2}{*}{No} & \multirow{2}{*}{8} & \multirow{2}{*}{Discrete} & \multirow{2}{*}{Dynamic} & \multirow{2}{*}{Arm} & \multirow{2}{*}{No} \\
    & & & Screen \\

    \midrule
    \cite{Lee2013} & 2013 &  Accelerometer & - & - & - & Gesture Classification & No & 20 & Discrete & Dynamic & Hand & No \\
    
    \midrule
    \cite{murugappan2013} & 2013 &  RGB-D Camera & - & - & HCI & Gesture Recognition & No & 3 & Continuous Discrete & Dynamic Static & Hand Fingers & Yes \\

    \midrule
    \cite{Zhou2014} & 2014 &  Accelerometer & Camera & - & - & Gesture Classification & No & 10 & Discrete & Dynamic & Hand & No \\

    \midrule
    \multirow{3}{*}{\cite{Iengo2014}} & \multirow{3}{*}{2014} &  \multirow{3}{*}{RGB-D Camera} & \multirow{3}{*}{-} & \multirow{3}{*}{-} & - & Gesture Recognition & No & 5 & Discrete & Dynamic & Arm & No \\
    & & & & & - & Gesture Recognition & No & 12 & Discrete & Dynamic & Full Body & No\\
    & & & & & HRI & Gesture Recognition & No & 3 & Discrete & Dynamic & Arm & No\\

    \midrule
    \multirow{2}{*}{\cite{Lu2014}} & \multirow{2}{*}{2014} &  \multirow{2}{*}{EMG} & \multirow{2}{*}{Accelerometer} & \multirow{2}{*}{-} & \multirow{2}{*}{HCI} & Gesture Recognition & No & 4 & Discrete & Static & Fingers Hand & No \\
    & & & & & & Gesture Recognition & No & 15 & Discrete & Static Dynamic & Fingers Arm & No \\

    \midrule
    \multirow{2}{*}{\cite{Lee2014}} & \multirow{2}{*}{2014} &  Stereo Dynamic & \multirow{2}{*}{-} & \multirow{2}{*}{-} & \multirow{2}{*}{-} & \multirow{2}{*}{Gesture Recognition} & \multirow{2}{*}{No} & \multirow{2}{*}{11} & \multirow{2}{*}{Discrete} & \multirow{2}{*}{Dynamic} & \multirow{2}{*}{Hand} & \multirow{2}{*}{No} \\
    & & Vision Sensor & & & & & \\

    \midrule
    \cite{KuiLiu2014} & 2014 &  RGB-D Camera & 6-Axis IMU & - & - & Gesture Classification & No & 5 & Discrete & Dynamic Static & Arm & No \\

    \midrule
    \cite{Yin2014} & 2014 &  Accelerometer & Button & - & - & Gesture Recognition & No & 8 & Discrete & Dynamic & Hand & No \\

    \midrule
    \cite{Ohn-Bar2014} & 2014 &  RGB-D Camera & - & - & - & Gesture Recognition & No & 19 & Discrete & Dynamic & Fingers Hand & No \\

    \midrule
    \cite{Duffner2014} & 2014 & 6-Axis IMU  & - & - & - & Gesture Classification & No & 9 & Discrete & Dynamic & Hand & No \\
    
    \midrule
    \cite{du2014} & 2014 & Infrared Camera  & - & - & HRI & Gesture Recognition & No & 2 & Continuous Discrete & Dynamic Static & Hand Fingers & No \\
    
    \midrule
    \cite{Xie2015} & 2015 & Accelerometer & - & - & - & Gesture Recognition & No & 12 & Discrete & Dynamic & Fingers & No \\

    \midrule
    \cite{Caramiaux2015} & 2015 & Infrared Camera & - & - & HCI & Gesture Recognition & No & 3 & Discrete & Dynamic & Hand & No \\

    \midrule
    \cite{Marques2015} & 2015 & Accelerometer & - & - & - & Gesture Classification & No & 7 & Discrete & Dynamic & Hand & No \\  
    
    \midrule
    \cite{Cicirelli2015} & 2015 & RGB-D Camera & - & - & HRI & Gesture Recognition & No & 10 & Discrete & Dynamic & Arm & No \\
    
    \midrule
    \multirow{2}{*}{\cite{Klboz2015}} & \multirow{2}{*}{2015} & Magnetic 3D & \multirow{2}{*}{-} & \multirow{2}{*}{-} & \multirow{2}{*}{HCI} & \multirow{2}{*}{Gesture Recognition} & \multirow{2}{*}{No} & \multirow{2}{*}{11} & \multirow{2}{*}{Discrete} & \multirow{2}{*}{Dynamic} & \multirow{2}{*}{Hand} & \multirow{2}{*}{No} \\
    & & Position Tracker\\
    
    \midrule
    \multirow{3}{*}{\cite{Yu-LiangHsu2015}} & \multirow{3}{*}{2015} & \multirow{3}{*}{9-Axis IMU} & \multirow{3}{*}{-} & \multirow{3}{*}{-} & \multirow{3}{*}{-} & Gesture Recognition & No & 10 & Discrete & Dynamic & Hand & No \\
    & & & & & & Gesture Recognition & No & 26 & Discrete & Dynamic & Hand & No \\
    & & & & & & Gesture Recognition & No & 8 & Discrete & Dynamic & Hand & No \\
    
    \end{tabular}}
\end{table*}

\begin{table*}[!ht]
    \ContinuedFloat
    \caption[Gestural literature classification (table 2/2).]{Classification of literature related to gesture-based interfaces - continued.}
    \centering
    \label{table:GestureClassification2}
    \resizebox{\textwidth}{!}{\begin{tabular}{c c c c c c c c c c c c c}
    \multicolumn{2}{c}{\textit{Article}} & \multicolumn{3}{c}{\textit{Sensing}} & \multicolumn{1}{c}{\textit{Reaction}} & \multicolumn{2}{c}{\textit{Processing}} & \multicolumn{5}{c}{\textit{Gestures}}\\
    \textit{Ref} & \textit{Year} &  \textit{Sensor (1)} & \textit{Sensor (2)} & \textit{Sensor (3)} & & \textit{Problem} & \textit{User defined} & \textit{Size} & \textit{Effect} & \textit{Time} & \textit{Focus} & \textit{Space}\\  
    \toprule
    
    \cite{Georgi2015} & 2015 & EMG & 6-Axis IMU & Button & - & Gesture Recognition & No & 12 & Discrete & Dynamic & Fingers Hand & No \\
    
    \midrule
    \cite{Wang2015} & 2015 & RGB-D Camera & - & - & - & Gesture Classification & No & 10 & Discrete & Static & Fingers & No \\
    
    \midrule
    \multirow{2}{*}{\cite{ramani2015}} & \multirow{2}{*}{2015} & RGB-D Camera & \multirow{2}{*}{-} & \multirow{2}{*}{-} & \multirow{2}{*}{HCI} & \multirow{2}{*}{Feature Extraction} & \multirow{2}{*}{No} & \multirow{2}{*}{-} & \multirow{2}{*}{Continuous} & \multirow{2}{*}{Dynamic} & \multirow{2}{*}{Hand Fingers} & \multirow{2}{*}{Yes} \\
    & & Infrared Camera& & & & & & & & & &  \\
    
    \midrule
    \cite{li2015} & 2015 & 9-axis IMU & - & - & - & Gesture Recognition & No & 11 & Discrete & Dynamic & Hand & No \\
    
    \midrule
    \cite{ahmad2015} & 2015 & Infrared Camera & - & - & - & Feature Extraction & No & 1 & Discrete & Dynamic & Fingers & Yes \\

    \midrule
    \cite{Galka2016} & 2016 & 7 Accelerometers & - & - & - & Gesture Classification & No & 40 & Discrete & Dynamic & Fingers Hand Arm & No \\

    \midrule
    \multirow{2}{*}{\cite{Moazen2016}} & \multirow{2}{*}{2016} & \multirow{2}{*}{Accelerometer} & \multirow{2}{*}{-} & \multirow{2}{*}{-} & \multirow{2}{*}{-} & Gesture Recognition & No & 5 & Discrete & Dynamic & Arm & No \\
    & & & & & & Gesture Recognition & No & 5 & Discrete & Dynamic & Arm & No \\
    
    \midrule
    \multirow{2}{*}{\cite{Hong2016}} & \multirow{2}{*}{2016} &  \multirow{2}{*}{Accelerometer} & \multirow{2}{*}{Button} & \multirow{2}{*}{-} & \multirow{2}{*}{-} & Gesture Recognition & No & 9 & Discrete & Dynamic & Hand & No \\
    & & & & & & Gesture Recognition & No & 7 & Discrete & Dynamic & Hand & No \\

    \midrule
    \cite{Wen2016} & 2016 & 6-Axis IMU & - & - & - & Gesture Recognition & No & 5 & Discrete & Dynamic & Fingers & No \\

    \midrule
    \multirow{2}{*}{\cite{Xie2016}} & \multirow{2}{*}{2016} &  \multirow{2}{*}{Accelerometer} & \multirow{2}{*}{Button} & \multirow{2}{*}{-} & \multirow{2}{*}{-} & Gesture Recognition & No & 8 & Discrete & Dynamic & Hand & No \\
    & & & & & & Gesture Recognition & No & 16 & Discrete & Dynamic & Hand & No \\

    \midrule
    \cite{Gupta2016continuous} & 2016 & 6-Axis IMU & - & - & HCI & Gesture Recognition & No & 6 & Discrete & Dynamic & Hand & No \\  

    \midrule
    \multirow{2}{*}{\cite{Gupta2016towards}} & \multirow{2}{*}{2016} & \multirow{2}{*}{RGB-D Camera} & Stereo & \multirow{2}{*}{-} & \multirow{2}{*}{-} & \multirow{2}{*}{Gesture Classification} & \multirow{2}{*}{No} & \multirow{2}{*}{25} & \multirow{2}{*}{Discrete} & \multirow{2}{*}{Dynamic} & \multirow{2}{*}{Fingers Hand} & \multirow{2}{*}{No} \\
    & & & Infrared Camera\\

    \midrule
    \multirow{2}{*}{\cite{Lai2016}} & \multirow{2}{*}{2016} &  \multirow{2}{*}{RGB-D Camera} & \multirow{2}{*}{-} & \multirow{2}{*}{-} & \multirow{2}{*}{HRI} & Gesture Recognition & No & 1 & Discrete & Static & (Bi) Arm & Yes \\
    & &  & & & & Feature Extraction & No & - & Continuous & Dynamic & Arm & No \\

    \midrule
    \multirow{2}{*}{\cite{Molchanov2016}} & \multirow{2}{*}{2016} & \multirow{2}{*}{RGB-D Camera} & Stereo & \multirow{2}{*}{-} & \multirow{2}{*}{HCI} & \multirow{2}{*}{Gesture Recognition} & \multirow{2}{*}{No} & \multirow{2}{*}{25} & \multirow{2}{*}{Discrete} & \multirow{2}{*}{Dynamic} & \multirow{2}{*}{Fingers Hand} & \multirow{2}{*}{No} \\
    & & & Infrared Camera\\
    
    \midrule
    \cite{zhou2016} & 2016 & RGB Camera & - & - & - & Gesture Classification & No & 14 & Discrete & Static & Hand Fingers & No \\

    \midrule
    \multirow{2}{*}{\cite{Haria2017}} & \multirow{2}{*}{2017} &  \multirow{2}{*}{RGB Camera} & \multirow{2}{*}{-} & \multirow{2}{*}{-} & \multirow{2}{*}{HCI} & Gesture Recognition & No & 6 & Discrete & Static & Fingers & No \\
    & &  & & & & Gesture Recognition & No & 1 & Discrete & Dynamic & Hand & No \\

    \midrule
    \cite{Coronado2017} & 2017 & Accelerometer & - & - & HRI & Feature Extraction & No & - & Continuous & Dynamic & Arm & No \\

    \midrule
    \multirow{2}{*}{\cite{Shin2017}} & \multirow{2}{*}{2017} & Epidermal & \multirow{2}{*}{-} & \multirow{2}{*}{-} & \multirow{2}{*}{-} & \multirow{2}{*}{Gesture Classification} & \multirow{2}{*}{No} & \multirow{2}{*}{5} & \multirow{2}{*}{Discrete} & \multirow{2}{*}{Static} & \multirow{2}{*}{Fingers Hand} & \multirow{2}{*}{No} \\
    & & Tactile Sensor\\

    \midrule
    \multirow{2}{*}{\cite{Mendes2017}} & \multirow{2}{*}{2017} & \multirow{2}{*}{Infrared Camera} & \multirow{2}{*}{-} & \multirow{2}{*}{-} & \multirow{2}{*}{HRI} & Gesture Recognition & No & 12 & Discrete & Static & Fingers Hand & No \\
    & &  & & & & Gesture Recognition & No & 10 & Discrete & Dynamic & Hand & No \\

    \midrule
    \cite{Naguri2017} & 2017 & Infrared Camera & - & - & - & Gesture Recognition & No & 6 & Discrete & Dynamic & Fingers Hand & No \\

    \midrule
    \cite{Oyedotun2017} & 2017 & RGB Camera & - & - & - & Gesture Classification & No & 24 & Discrete & Static & Fingers Hand & No \\

    \midrule
    \cite{Bao2017} & 2017 & RGB Camera & - & - & - & Gesture Classification & No & 7 & Discrete & Static & Fingers & No \\
    
    \midrule
    \cite{Liang2017} & 2017 & RGB-D Camera & - & - & - & Feature Extraction & No & - & Continuous & Dynamic & Hand & Yes \\

    \midrule
    \cite{Kim2018} & 2018 & 6-Axis IMU & - & - & - & Gesture Recognition & No & 9 & Discrete & Dynamic & Arm & No \\

    \midrule
    \cite{Bolano2018} & 2018 & RGB-D Camera & - & -  & HRI & Feature Extraction & No & - & Continuous & Dynamic & Hand & Yes \\

    \midrule
    \cite{Buddhikot2018} & 2018 & RGB Camera & - & - & HCI & Gesture Recognition & No & 6 & Discrete & Dynamic & Fingers Hand & No \\

    \midrule
    \multirow{2}{*}{\cite{Zeng2018}} & \multirow{2}{*}{2018} & \multirow{2}{*}{Infrared Camera} & \multirow{2}{*}{-} & \multirow{2}{*}{-} & \multirow{2}{*}{-} & Gesture Classification & No & 10 & Discrete & Dynamic & Fingers & No \\
    & &  & & & & Gesture Classification & No & 26 & Discrete & Dynamic & Fingers & No \\

    \midrule
    \cite{Hu2018} & 2018 & RGB Camera & - & - & - & Gesture Recognition & No & 5 & Discrete & Dynamic & Fingers Arm & No \\

    \midrule
    \cite{Ma2018} & 2018 & RGB-D Camera & - & - & - & Gesture Recognition & No & 8 & Discrete & Dynamic & (Bi) Hand Arm & No \\
    
    \midrule
    \cite{Carfi2018} & 2018 & Accelerometer & - & - & - & Gesture Recognition & No & 6 & Discrete & Dynamic & Arm & No \\
    
    \midrule
    \cite{Kim2019} & 2019 & Accelerometer & Button & - & - & Gesture Recognition & No & 10 & Discrete & Dynamic & Hand & No \\

    \midrule
    \cite{Islam2019} & 2019 & RGB Camera & - & - & HRI & Gesture Recognition & No & 8 & Discrete & Static & Fingers & No \\

    \midrule
    \multirow{5}{*}{\cite{Huang2019}} & \multirow{5}{*}{2019} & \multirow{5}{*}{RGB-D Camera} & \multirow{5}{*}{-} & \multirow{5}{*}{} & - & Gesture Recognition & No & 10 & Discrete & Dynamic & Hand Arm & No \\
    & & & & & - & Gesture Recognition & No & 10 & Discrete & Dynamic & Hand Arm & No \\
    & & & & & HCI & Gesture Recognition & No & 10 & Discrete & Dynamic & Hand Arm & No \\
    & & & & & HCI & Gesture Recognition & No & 10 & Discrete & Dynamic & Hand Arm & No \\
    & & & & & HCI & Gesture Recognition & No & 15 & Discrete Continuous & Static Dynamic & Fingers Arm & No \\

    \midrule
    \multirow{2}{*}{\cite{Park2019}} & \multirow{2}{*}{2019} & \multirow{2}{*}{Infrared Camera} & \multirow{2}{*}{-} & \multirow{2}{*}{-} & HRI & Feature Extraction & No & - & Continuous & Dynamic & Hand & No \\
    & & & & & HCI & Gesture Recognition & No & 1 & Discrete Continuous & Static Dynamic & Fingers Hand & Yes \\

    \midrule
    \cite{Pomboza-Junez2019} & 2019 & EMG & - & - & HCI & Gesture Recognition & No & 4 & Discrete & Static & Fingers Hand & No \\

    \midrule
    \multirow{2}{*}{\cite{Avola2019}} & \multirow{2}{*}{2019} & Stereo & \multirow{2}{*}{-} & \multirow{2}{*}{-} & \multirow{2}{*}{-} & \multirow{2}{*}{Gesture Classification} & \multirow{2}{*}{No} & \multirow{2}{*}{30} & \multirow{2}{*}{Discrete} & \multirow{2}{*}{Static Dynamic} & \multirow{2}{*}{Fingers Hand} & \multirow{2}{*}{No} \\
    & & Infrared Camera\\

    \midrule
    \multirow{2}{*}{\cite{Feng2019}} & \multirow{2}{*}{2019} & Stereo & \multirow{2}{*}{-} & \multirow{2}{*}{-} & \multirow{2}{*}{-} & \multirow{2}{*}{Gesture Classification} & \multirow{2}{*}{No} & \multirow{2}{*}{10} & \multirow{2}{*}{Discrete} & \multirow{2}{*}{Static} & \multirow{2}{*}{Fingers} & \multirow{2}{*}{No} \\
    & & Infrared Camera\\

    \midrule
    \cite{Zhang2019} & 2019 & EMG & - & - & - & Gesture Recognition & No & 5 & Discrete & Dynamic & Fingers Hand & No\\

    \end{tabular}}
\end{table*}    

\section{Literature classification}
\label{sec:classification}
As we observed above, the literature about gesture-based interaction is vast and heterogeneous.
In this Section, we exploit the conceptual tools developed in the previous Sections to analyse and classify it.
It is important to note that it should not be considered as a systematic nor in-depth analysis of the literature.
Instead, it should be treated as an example of how literature can be classified adopting an operational perspective using the taxonomy we introduce in this paper.
Table \ref{table:GestureClassification} includes all the reviewed articles classified using the taxonomy introduced in Section \ref{sec:2:taxonomy}. 
Articles are ordered chronologically.
All the columns of the Table are grouped into five categories:

\begin{itemize}
    \item \textit{Article info} includes two sub-columns, namely the article reference and the publication year.
    \item \textit{Sensing} focuses on the employed sensing modalities and reports multiple sensors whenever they are used.
    \item \textit{Reaction} describes generically whether a certain work refers to HCI or HRI studies, i.e., whether the reaction involves a virtual or a physical system. Here we consider only those papers whereby such reaction is clearly described, or a validation is presented. Articles not fulfilling these requirements are classified as not having a system reaction.
    \item \textit{Processing} includes two sub-columns, the former with the specification of the problem actually solved, i.e., recognition, feature extraction or classification, the latter with information about user-defined gestures.
    \item \textit{Gestures} is the proper gestures classification, including the size of the dictionary for discrete gestures, as well as the effect, time, focus and spatial characteristics.
\end{itemize}
In the Table, the articles which consider distinctly discrete and continuous gestures are described using two rows for the gesture and processing groups. 
This happens even when an article elaborates on multiple dictionaries.
If different gestures have different application scenarios, an extra row is added to the reaction column as well. 
The focus column can refer to more than one body part, on the basis of the specific article contents.
The main factors we considered to determine the focus of the gestures related to a specific work are the description of gestures, pictures, or videos whereby gestures are shown, employed sensors and computational approaches. 
In the Table, when the focus-related column contains multiple entries, while the columns related to effect and time contain one entry only, the latter characterise the gesture of each of the referred body parts, an example being the row in \cite{Yeasin2000} for a discrete, dynamic, finger/hand gesture.
If the time column contains both the static and dynamic keywords, and just one body part is specified in the focus column, then the dictionary contains both static and dynamic gestures performed with that body part, e.g. in \cite{Kim2007} one discrete, static, hand gesture and four discrete, dynamic, hand gestures are used. 
If both the focus and time columns contain two entries, this means that the dictionary is composed of gestures whereby one body part is static and the other is dynamic.
The order is preserved in-between columns, e.g., in \cite{Avola2019} are used discrete, static, fingers as well as dynamic hand gestures. 
This, of course, can be extended even to the case whereby effect, time and focus columns contain two keywords, e.g., in \cite{Starner2000} discrete, static, fingers and continuous, dynamic, arm gestures.
The \textit{bi} tag has been added to the focus column in all cases addressing the processing problem simultaneously for different body parts. 
Works where bi-manual gestures are considered as the combination of two single limb gestures combined after the processing phase, as done for instance in \cite{Islam2019}, are not identified by this tag.

While performing the analysis of relevant literature, we have encountered a huge variety of meanings associated with the gesture recognition problem. 
Often, the term ``gesture recognition'' is used to refer to whichever procedure related to gesture-based sensing and processing. 
Nevertheless, and to the best of our knowledge, we did not encounter works that we could not represent using our taxonomy. 
Notice that, since our study focuses on functional HMI only articles considering intentional gestures have been included in the analysis. Furthermore, a few articles have been discarded because of their lack of clarity in presentation.
However, the idea of the authors is to keep this list updated on a dedicated website\footnote{\label{website}Web: \url{https://acarfi.github.io/GesturalInteractionSurvey}} accepting suggestions by the community to enrich it.

More extensive and methodical reviews exist, which may be useful to perform relevant statistical analyses on the kind of adopted sensors, gesture types, and dictionary size. 
Therefore, we do not want to perform this kind of observations here.
However, it is noteworthy to point out one single observation related to how our taxonomy can be used. 
According to the problem statement, in order for the interface to react to a gesture-related stimulus, the processing phase should be completed. 
As a consequence, the problems to be solved are feature extraction for continuous gestures, and gesture recognition for discrete gestures. Consistently, in Table \ref{table:GestureClassification} each work that exhibits a reaction solves either feature extraction or gesture recognition.

\begin{table*}[!ht]
    \caption[Discrete gesture literature.]{Summary of literature for discrete dynamic arm gestures spatially unrelated perceived using IMUs.}
    \centering
    \label{table:IMUstateoftheart}
    \resizebox{\textwidth}{!}{\begin{tabular}{c c c c c c c c c c c c c c c}
    \toprule
    \multicolumn{2}{c}{Article info} & Sensing & Reaction & \multicolumn{2}{c}{Gestures} & Frequency & (0) Preprocessing & (1) Feature & (2) Detection & (3) Classification & \multicolumn{4}{c}{Order}\\
    Ref & Year &  &  & Size & Focus & & & Extraction & & \\  
    \toprule

    \cite{Xing-HanWu2010} & 2010 &  Accelerometer & HRI & 6 & Arm & 1600 Hz & Yes & No & Threshold & DTW & 2 & 0 & 3 & -\\
    
    \midrule
    \cite{Xie2015} & 2015 & Accelerometer & - & 12 & Fingers & 50 Hz & Yes & Yes & Threshold & Threshold & 0 & 2 & 1 & 3\\
    
    \midrule
    \multirow{3}{*}{\cite{Yu-LiangHsu2015}} & \multirow{3}{*}{2015} & \multirow{3}{*}{9-Axis IMU} & \multirow{3}{*}{-} & 10 & Hand & \multirow{3}{*}{75} & \multirow{3}{*}{Yes} & \multirow{3}{*}{Yes} & \multirow{3}{*}{Threshold} & \multirow{3}{*}{DTW} & \multirow{3}{*}{0} & \multirow{3}{*}{2} & \multirow{3}{*}{1} & \multirow{3}{*}{3} \\
    & & & & 26 & Hand \\
    & & & & 8 & Hand \\

    \midrule
    \multirow{2}{*}{\cite{Moazen2016}} & \multirow{2}{*}{2016} & \multirow{2}{*}{Accelerometer} & \multirow{2}{*}{-} & 5 & Arm & \multirow{2}{*}{-} & \multirow{2}{*}{Yes} & \multirow{2}{*}{Yes} & \multirow{2}{*}{Threshold} & \multirow{2}{*}{DTW} & \multirow{2}{*}{0} & \multirow{2}{*}{1} & \multirow{2}{*}{2} & \multirow{2}{*}{3}\\
    & & & & 5 & Arm \\

    \midrule
    \multirow{4}{*}{\cite{Wen2016}} & \multirow{4}{*}{2016} & \multirow{4}{*}{6-Axis IMU} & \multirow{4}{*}{-} & \multirow{4}{*}{5} & \multirow{4}{*}{Fingers} & \multirow{4}{*}{50} & \multirow{4}{*}{Yes} & \multirow{4}{*}{Yes} & & SVM & \multirow{4}{*}{0} & \multirow{4}{*}{1} & \multirow{4}{*}{2}& \multirow{4}{*}{3}\\
    & & & & & & & & & \multirow{2}{*}{DTW-k-NN} & Naive Bayes \\
    & & & & & & & & & \multirow{2}{*}{+ Threshold} & Logistic Regression \\
    & & & & & & & & & & k-NN \\

    \midrule
    \cite{Gupta2016continuous} & 2016 & 6-Axis IMU & HCI & 6 & Hand & - & Yes & Yes & Threshold & DTW & 0 & 1 & 2 & 3\\

    \midrule
    \cite{Kim2018} & 2018 & 6-Axis IMU & - & 9 & Arm & - & Yes & No & Threshold & GRU NN & 2 & 0 & 3 & -\\

    \bottomrule
    \end{tabular}}
\end{table*}

\section{An Example of Usage}
\label{sec:eou}

This Section presents and briefly discusses one example of how the literature analysis introduced above, if combined with the gesture taxonomy and the GI-UI problem statement, can help researchers and HMI designers build effective systems on top of the current state-of-the-art. 

Let us hypothesise that a researcher or designer is interested in an HMI system 
(i) employing discrete, dynamic arm gestures that are not spatially related, and 
(ii) that the application under design requires the usage of IMUs. Given these functional requirements, it is possible to perform a selection of the works considered in the literature analysis, and therefore understanding the relationships between employed sensors, data processing and system reactions that are present in the literature. 

For the specific example, such a selection is reported in Table \ref{table:IMUstateoftheart}, where the common features of the identified works in the literature have been removed in favour of a more detailed description of the GI-UI problem solutions. 
In particular, columns have been added to describe the approach used to solve the gesture recognition problem, namely frequency of data acquisition, presence of a pre-processing step, presence of a feature extraction step, detection technique, classification technique, and the order in which the four phases are executed.

From the Table, the advantages of the taxonomy become evident.
Starting with an idea about target gestures and the sensors to be employed, it is possible to get a complete overview of typically employed body parts, data acquisition frequency, data processing pipeline, as well as the algorithms employed for classification in state-of-the-art approaches.

\section{Discussion}
\label{sec:discussion}

In Section \ref{sec:data_processing}, we have described how data processing for continuous and discrete gestures implies facing two different problems, i.e., feature extraction and gesture recognition, respectively. 
Furthermore, gesture recognition has been organised as a pipeline involving three computational steps, namely pre-processing and feature extraction, gesture detection, and gesture classification. 
In the literature analysis we carried out, this characterisation has been used to identify the specific problems each referenced work aimed at addressing. 
If we refer to Table \ref{table:GestureClassification}, contributions in the literature focusing on the gesture recognition problem are expected to address all these three computational steps. 
Instead, the ones addressing gesture classification focus mainly on techniques to classify gestures, although usually they include and employ procedures for pre-processing and feature extraction as well.
In all these articles, and often without a clear statement of purpose nor an explicitly stated assumption, the authors rely on the closed-world assumption, i.e., the presumption that the system has complete knowledge about all possible interactions, and whichever action is performed by the user it is part of the gesture dictionary. 
In order to use these classification techniques in real-world scenarios, the closed-world assumption should be relaxed, and therefore an open-world approach should be considered, i.e., the idea that the system does not have complete knowledge, and therefore actions or gestures performed by the user may be unknown.

A shift towards an open-world assumption is possible if a suitable detection procedure is employed. 
Although its importance for real-world applications is evident, and notwithstanding the vast attention it received in the context of data-driven approaches, where it is often referred to as \textit{novelty detection} \cite{pimentel2014review}, the detection procedure does not attract the interests of researchers working on gesture-based interaction, whereas the main focus remains the development of new techniques to solve the classification problem using data-driven approaches. 
One reason for that lack of attention may be related to the difficulty associated with the evaluation of new classification metrics with respect to state-of-the-art approaches, especially when compared to a similar evaluation in case of gesture detection, i.e., benchmarking  \cite{ward2011performance}.
Therefore, gesture detection approaches are often simplified, i.e., using threshold-based mechanisms, or even naive methods, such as asking users to hold a button while performing a gesture to simplify data segmentation.
We argue that future works in discrete gestural interaction should prioritise studies focusing on the \textit{overall gesture recognition} and in particular on its \textit{detection} aspects.

It is noteworthy that continuous gestures are under-represented in our analysis. This could be the result of a bias in the analysis methodology. 
However, our findings are consistent with the ones of a recent systematic literature review where 89\% of the reviewed articles consider discrete gestures \cite{Vuletic2019}. 
Nonetheless, a number of interesting considerations regarding continuous gestures can be done. 
Researchers interested in continuous gestures \textit{should not completely disregard} the literature focused on discrete ones since, in order to solve the gesture recognition problem, feature extraction should be addressed as well.
In fact, many articles focusing on discrete gestures include interesting feature extraction analyses, which could be applied to continuous gestures \cite{zhou2016}. 

Studies focusing only on continuous gestures are usually limited to reactions that are conceptually simple, e.g., the control of a mobile robot \cite{Coronado2017}, or a \textit{one to one} mapping between the human hand and the robot end-effector for tele-operation purposes \cite{Bolano2018}, where the main problem to solve is geometric in nature.
In order to attain more complex reactions, future works in the field should consider \textit{a principled integration} between continuous and discrete gestures. 
One possible example is a GUI using continuous gestures to move a cursor and adopting at the same time discrete gestures to implement icon selection.
Therefore, we believe that novel studies to address the GI-UI problem should focus on the integration of state-of-the-art solutions for continuous and discrete gestures.

Finally, it is remarkable that research on the GI-UI problem mainly focuses on upper limb gestures. 
This is of course motivated by the intuitiveness in interacting with physical or virtual systems using hands. 
However, in daily life scenarios, humans extensively use hands to interact with the environment. 
Therefore, future works should extend gesture-based interaction to other body parts, e.g., feet. 
This could unlock new interaction modalities even in situations whereby hands are already occupied, with obvious positive consequences in case of specialised interfaces for people with special needs.

\section{Conclusions}

In this paper, we have analysed relevant state-of-the-art approaches for gestural interaction in a broad sense but focusing on functional human-machine interaction.

Our analysis starts from the observation, supported by previous research \cite{Vuletic2019}, that two essential aspects are missing from the literature: a shared vocabulary and a clear problem statement.
Therefore, we have first tried to structure the problem giving a formal definition of what a functional gesture is, and defining a taxonomy considering four gesture characteristics, i.e., effect, time, focus and space. 
Then, we have formalised the Gestural Interaction for User Interfaces (GI-UI) problem, which is structured on three sub-problems, specifically related to sensing, data processing and system reaction, and we have described how each characteristic influences the problem statement and the technique adopted for its solution.

In the paper we have classified $83$ articles using the proposed gesture taxonomy according to the formalised problem statement. 
The classification has the goal of helping researchers to look for contributions addressing a GI-UI problem independently of the application field. 
The classification is published on a dedicated website
, and the authors have the intention to update it accepting the help of the community. 

On top of these results, it is possible to design and develop new gesture-based methodologies and technologies with the aim of taking advantage of a full integration of continuous and discrete gestures.
As a next step, the conceptualisation efforts carried out in this study will be extended to tackle related problems in human motion interpretation, with the aim of moving towards a general representation of human motion.


\ifCLASSOPTIONcaptionsoff
  \newpage
\fi

\bibliographystyle{IEEEtran} 
\bibliography{overall}


%







\end{document}